\documentclass[manuscript]{aastex}
\usepackage{graphicx}
\usepackage{bm}
\def\alf{Alfv\'en\,}
\def\bq{\begin{equation}}
\def\eq{\end{equation}}
\let\grad=\nabla

\def\drho{\delta\rho}
\def\dP{\delta P}

\def\v{{{\bf{v}}}}
\def\u{{{\bf{u}}}}
\def\duz{\delta u_z}
\def\dux{\delta u_x}
\def\duy{\delta u_y}
\def\dup{\delta u_+}
\def\dum{\delta u_-}

\def\vi{{{\bf{v}}}_i}
\def\vj{{{\bf{v}}}_j}
\def\ve{{{\bf{v}}}_e}
\def\vi{{{\bf{v}}}_i}
\def\vd{{{\bf{v}}}_d}

\def\vn{{{\bf{v}}}_n}
\def\B{{\bf{B}}}
\def\dBx{\delta B_x}
\def\dBy{\delta B_y}
\def\dBz{\delta B_z}
\def\dBp{\delta B_+}
\def\dBm{\delta B_-}

\def\dB{{\delta{\bf\B}}}
\def\J{{{\bf{J}}}}
\def\E{{{\bf{E}}}}

\def\du{{\bf{\delta\u}}}

\def\dB{{\bf{\delta\B}}}
\def\dJ{{\bf{\delta\J}}}

\def\sigv{<\sigma v>}
\newcommand\cross{{\bf{\times}}}
\def\curl{{\grad \cross}}
\newcommand{\dt}[1]{\frac{d #1}{dt}}
\newcommand{\delt} [1] {\frac{\partial #1}{\partial t}}
\newcommand{\delz} [1] {\frac{\partial #1}{\partial z}}
\shorttitle{ Parametric instability in molecular clouds}
\shortauthors{Pandey $\&$ Vladimirov}

\begin{document}
\title{ Parametric instability in dark molecular clouds}
\author{B. P. Pandey}
\affil{Department of Physics, Macquarie University, Sydney, NSW 2109,
Australia}
\email{bpandey@physics.mq.edu.au}

\and

\author{S. V. Vladimirov}
\affil{ School of Physics, The University of Sydney, NSW 2006, Australia}
\email{S.Vladimirov@physics.usyd.edu.au}


%
%
\begin{abstract}
The present work investigates the parametric instability of parallel propagating circularly polarized \alf (pump) waves in a weakly ionized molecular cloud. It is shown that the relative drift between the plasma particles gives rise to the Hall effect resulting in the modified pump wave characteristics. Although the linearized fluid equations with periodic coefficients are difficult to solve analytically, it is shown that a linear transformation can remove the periodic dependence. The resulting linearized equations with constant coefficients are used to derive an algebraic dispersion relation. The growth rate of the parametric instability is a sensitive function of the amplitude of the pump wave as well as to the ratio of the pump and the modified dust-cyclotron frequencies. The instability is insensitive to the plasma $\beta$. The results are applied to the molecular clouds.
\end{abstract}

\section{INTRODUCTION}

Molecular clouds are thought to be the nursery of star formation \citep{l85}. These clouds generally consist of very small ($\sim 10^{-4} -– 10^{-8}$) fraction of ionized matter and yet, depending upon the level of plasma-neutral coupling, the magnetic field can play an important role in the dynamics of such a cloud. Thus the question of how well the magnetic field is coupled to the neutral matter becomes crucial for various models of cloud support and star formation. The collisonal momentum exchange and the ratio of the gyro to various collision frequencies determine this coupling. The relative drift between the plasma particles introduces another interesting feature to this complex scenario. For example, when ions and neutrals are strongly coupled, the relative drift of the ``frozen-in" ions against the sea of neutrals may cause the diffusion of the magnetic flux reducing the flux to mass ratio in the star forming regions 
\citep{ms56, mtk06}. When the ions and neutrals are moving together and there is relative drift between the electrons and ions, the redistribution of energy and angular momentum in the medium may take place \citep{w99}. Thus, in molecular clouds, investigation of the non-ideal magnetohydrodynamic (MHD) effects is important. There are three regimes in which non-ideal MHD effects may operate, (I) ambipolar - when the magnetic field can be regarded as frozen in the plasma and drifts with it through the neutrals (applicable to the relatively low density, high ionization fraction regions), (II) Ohmic - when neutrals stops the ionized particle from drifting with the field (applicable to the high density, low ionization fraction regions) and (III) Hall – when electrons are well coupled or partially coupled to the field and ions and grains are partially or completely decoupled from the field.

The magnetic field can provide either static or dynamic (by magnetohydrodynamic waves) support to the molecular clouds \citep{d70, ms76, sal87, mhp97, fp00}. The \alf waves are not easily dissipated \citep{zj83} making them an attractive candidate for the dynamic support of the cloud. The origin of the non-thermal line broadening in molecular cloud is attributed to the Alfvenic turbulence owing to the non-steepening of the large amplitude \alf waves \citep{fh02, fhw04, mtk06}. The supernova shocks, gravitational collapse or the large scale motion of the cloud could act as the source of such waves in the medium.

In dense molecular cloud cores charged grains are more numerous in number than the electrons and ions \citep{wn99}.  The ionization fraction of the plasma is strongly affected by the abundance and size distribution of the grains through the recombination process on the grain surface. In molecular clouds with densities $n_n \geq 10^3\,\mbox{cm}^{-3}$, inelastic collision of various charged species with the grain is important \citep{cm93}. At the densities relevant to cloud cores and at higher densities occurring during the formation of protostar and protostellar discs (densities $\geq 10^7 - 10^{11}$ cm$^{- 3}$), the role of the grains can not be neglected in the dynamics \citep{w99}. The presence of the charged grains in the planetary and interstellar medium causes the excitation of very low frequency \alf waves \citep{ph87, cv94, dej05}. These waves provide important physical mechanism for the transport of angular momentum and energy in a differentially rotating dusty disks \citep{w99, pw06}.

The parametric instability of a finite amplitude circularly polarized \alf wave have been studied for last several decades \citep{d78, g78, h94}. Recently, this problem has been studied in the single fluid MHD framework for a self-gravitating molecular cloud \citep{fh99}. We shall leave the Jeans instability out of the present investigation and revisit the problem of parametric instability in the molecular clouds by including the charged grains and neutrals in the formulation. The investigation of parametric instability in a dusty medium is relatively recent phenomena \citep{vos05, hcv03}. The charged grains are assumed immobile and the pump wave is assumed magnetosonic or Alfvenic. A recent generalization of the parametric investigation to the multi-component plasma \citep{hcv04} removes the limitation of the immobile dust grains. However, the dynamics of the neutrals has been left out in making it inapplicable to the interstellar medium.

It is known that collisonal processes in weakly ionized plasma can strongly influence the nonlinear ambipolar diffusion and the ensuing current sheet formation \citep{cv94}.  The role of the plasma-dust collision may as well be important for the parametric instability. For example, collisions between the plasma particles and grains are responsible for some of the novel features in dusty plasma. The new collective behavior is known to exist in such plasma due to the charge fluctuations -- an offshoot of collision \citep{vos05, bp94}. Therefore, it is desirable to investigate the propagation of the large amplitude \alf wave in a weakly ionized collisonal plasma.  We note that in dark molecular clouds, the momentum of the bulk fluid is carried by the neutrals and the current is carried by the ionized component. As we shall see this will provide some economy of description to the present problem.

One of the difficulties in the investigation of parametric instability of circularly polarized \alf wave is associated with the appearance of periodic coefficients in the linearized MHD equations. This restricts the normal mode analysis of the spatial dependence and thus, the resulting Floquet equation can only be solved in the long wavelength limit. However, a recently proposed linear transformation \citep{rs03} reduces the periodic coefficients in the linearized MHD equations to the constant coefficient. Although the transformation by \cite{rs03} was applied to the non-dispersive \alf wave, it is shown in the present work that such a transformation is equally applicable to a more general, dispersive \alf modes.

In the present work, it is assumed that the charge on dust grains is constant. The basic formulation is discussed in Sec. II. It is assumed that the electrons and ions are well coupled to the magnetic field and essentially move together. This will restrict the applicability of the result to the regions of molecular cloud where neutral number density $n_n \leq 3\,\times\,10^9\,\mbox{cm}^{-3}$ \citep{dm01, tm05}. Assuming such equality, the induction equation can conveniently be written in the neutral frame. The resulting induction equation is shown to contain the Hall term, due to the relative drift between the plasma and the dust grains. It is known that this feature is a generic property of the multi-component plasmas \citep{w99}. The \alf pump wave is discussed in Sec. III and linearized equations are derived. Applying a linear transformation, a set of equation with constant coefficient is derived. In Sec. IV the dispersion relation is numerically solved. The dependence of the parametric instability on the pump wave amplitude and on the ratio of the pump wave to the dust-cyclotron frequency is discussed. In Sec. V, application of the results to the molecular cloud is discussed, and conclusion and brief summary of the results are presented.

\section{FORMULATION}

We shall assume a weakly ionized molecular cloud consisting of plasma particles, i.e. electrons and ions, charged grains and neutral particles. The dynamics of such a partially ionized mixture is described by the multi-fluid system. The basic set of equations describing the dynamics of such a system is as follows. The continuity equation is 
\bq 
\frac{\partial \rho_j}{\partial t} + \grad\cdot\left(\rho_j\,\vj\right)=0\,. \label{eq:ccnt} 
\eq 
Here, $\rho_j$ is the mass density and $\vj$ is the velocity of the various plasma components and neutrals. We note that the plasma velocities $\vj$ are written in the neutral frame. The momentum equations for electrons, ions, grains and neutrals are 
\bq 
0 = - q_j\,n_j\,\left(\E' + \frac{\vj\cross \B}{c}\right) -
\rho_j\,\nu_{jn}\,\vj\,, 
\label{eq:em1} 
\eq
\bq \rho_n\,\dt\,\vn = - \grad{P} + \displaystyle\sum_{e, i,d}
\rho_j\,\nu_{j n}\,\vj\,. 
\label{eq:nm1} 
\eq 
Here $ q_j\,n_j\,\left(\E' + \vj\cross \B/c\right)$ is the Lorentz force and $\E' = \E + \vn\cross \B / c$ is the electric field in the neutral frame with $\E$ and $\B$ as the electric and magnetic fields respectively,  $n_j$ is the number density, and,  $j$ stands for electrons ($q_e = -e$), ions ($q_i = e$) and dust $q_d = Z\,e$ with $Z$ as the number of charge on the grain and $c$ is the speed of light.

The collision frequency is
\bq 
\nu_{j\, n} \equiv \gamma_{j\, n}\,\rho_n = \frac{\sigv_j}{m_n +
m_j}\,\rho_n\,. 
\label{cf0} 
\eq 
Here $\sigv_j$ is the rate coefficient for the momentum transfer by the collision of the $j^{\mbox{th}}$ particle with the neutrals.

The ion-neutral and electron-neutral rate coefficients are (Draine et al. 1983)
\begin{eqnarray}
<\sigma\,v>_{in} = 1.9 \cross 10^{-9}\quad
\mbox{cm}^3\,\mbox{s}^{-1}\nonumber \\
<\sigma\,v>_{en} = 4.5 \cross
10^{-9}\,\left(\frac{T}{30\,\mbox{K}}\right)^{\frac{1}{2}}\quad
\mbox{cm}^3\,\mbox{s}^{-1}.
\label{cf1}
\end{eqnarray}
Adopting a value of $m_i = 30\,m_p$ for ion mass and $m_n = 2.33\,m_p$ for mean neutral mass where $m_p = 1.67 \times 10^{-24}\,\mbox{g}$ is the proton mass, the ion neutral collision frequency can be written as 
\bq 
\nu_{in} = \rho_n\,\gamma_{in} \equiv \frac{m_n\,n_n\,<\sigma\,v>_{in}}{m_i + m_n} = 1.4\cross 10^{-10}\,n_n\,\mbox{s}^{-1}\,. 
\label{cf2} 
\eq 
For very small grains ($\sim 3 - 3000\,\AA$), the grain-neutral collision frequency is of the same order as the ion-neutral collision frequency, given by equation (\ref{cf2}). However, for larger grains, the grain-neutral collision rate can vary between $10^{-10}$ to $10^{-5}\,\mbox{s}^{-1}$ for sizes ranging between a few Angstrom to a few microns. This can be seen if we write the collision rate as \citep{nu86}
\begin{equation}
<\sigma\,v>_{dn} = 2.8 \cross 10^{-5}\, T_{30}^{\frac{1}{2}}\,a_{-5}^{2}\,,
\label{cg}
\end{equation}
where $T_{30}$ is the gas temperature and $a_{-5}$ is the grain radius in the units of 30\,K and $10^{-5}$\,cm respectively.

We shall define the mass density of the bulk fluid as $ \rho \approx \rho_n$. Then the bulk velocity is $ \u \approx \v_n$.  The continuity equation (summing up equation (\ref{eq:ccnt})) for the bulk fluid becomes 
\bq 
\delt\rho + \grad\cdot\left(\rho\,\u\right) = 0\,. 
\label{eq:cont1} 
\eq 
The momentum equation can be derived by adding equations (\ref{eq:em1}) and (\ref{eq:nm1})
 \bq
\rho\,\frac{d\u}{dt} =  - \nabla\,P + \frac{\J\cross\B}{c}\,.
\label{eq:meq} 
\eq

As noted above, we shall assume that the electrons and ions are well coupled to the magnetic field. Defining 
\bq
\beta_j = \frac{\omega_{cj}}{\nu_{jn}}\,,
\eq
as the ratio of cyclotron $\omega_{cj} = q_j\,B/m_j\,c$ to the collision $\nu_{jn}$ frequencies, we shall thus assume that $\beta_e \gg \beta_i \gg 1$, i.e. plasma particles are tied to the magnetic field and move together $\ve \simeq \vi $. Then making use of the plasma quasi-neutrality condition, $n_e = n_i + Z\,n_d$, expression for the current density $\J = e\,\left(n_i - n_e \right)\,\v_e + Z\,e\,n_d\,\vd$ yields
\bq 
\ve = \frac{- \J}{Z\,e\,n_d} + \vd\,. 
\label{eq:vdf} 
\eq
It is known that the grain size may vary between few microns down to few tens of atoms in the interstellar medium. In fact, the presence of very small grains like MRN distributions \citep{mrn77}, and polycyclic aromatic hydrocarbons \citep{plb85} can play an important role in the dynamics. Since small grains are numerous and their charge may vary between $\pm 1$ and $0$ \citep{nnu91}, grains could be coupled to the magnetic field either directly, i.e. $\beta_d \gg 1$ or indirectly, i.e. even when $\beta_d \ll 1$ but $(n_i/n_d)\,\beta_d\,\beta_i \geq 1$ \citep{dm01}. Following \cite{cm93}, we shall express grain velocity $\vd$ in terms of electron velocity as  
\bq
\vd = \frac{\Theta}{1 + \Theta} \ve\,,
\eq
where
\bq
\Theta = \left[1 + \frac{\nu_{nd}}{\nu_{ni}}\right]\,\beta_d^2\,.
\eq
Thus, equation (\ref{eq:vdf}) becomes
\bq
\ve =  \frac{- \left(1 + \Theta\right)}{Z\,e\,n_d}\,\J.
\eq

Taking curl of the electron momentum equation (\ref{eq:em1}) and making use of Maxwell's equation, in the $\beta_e \gg 1$ limit the induction equation can be written as
\bq 
\delt \B = \curl\,\left[\left(\u\cross\B\right) - \frac{\left(1 + \Theta\right)}{Z\,e\,n_d} \J\cross\B\right]\,. 
\label{eq:ind} 
\eq
We shall note that in the absence of dust, $\J\cross\B /Z\,e\,n_d \rightarrow 0$, and, Hall term will disappear in the induction Eq. (\ref{eq:ind}). The ideal MHD description of the two component plasma assumes that the relative drift between electrons and ions are absent, i.e., $\ve = \vi$. The reason for the Hall effect in such a plasma is due to the relative drift between the plasma particles, since ${\E + \ve\cross\B}/c = 0$, can be written as ${\E + \vi\cross\B}/c = {\J\cross\B}/e\,n_e$ in the presence of the drift. However, in the presence of charged grains, or for that matter in any multi-component plasma, since $n_e \neq n_i$, even when relative drift between the plasma particles are absent, i.e., $\ve = \vi$, owing to the presence of the third, charged component, Hall effect will always be present. This can be seen from Eq. (\ref{eq:vdf}). The Hall effect disappears in such a plasma if in addition to the absence of relative drift, one also demand $Z\,n_d \ll n_e$. In molecular clouds generally $Z\,n_d \leq n_e$  and thus, Hall drift is always present. Therefore, the Hall effect in a multi-component plasma can be caused either by the relative drift between the plasma particles or due to the presence of the charged grains. This is the reason why Hall MHD description of a multi-component plasma could be the proper description of the magnetized protoplanetary disks \citep{w99, bt01}.

The ratio of the Hall to the convective term in the induction equation (\ref{eq:ind}) will shed the light on the relative importance of the two terms and the role of the grain magnetization. The ratio of the two terms can be written as
\bq
\frac{c}{V_A}\frac{1+\Theta}{4\,\pi\,Z\,e\,n_d}{|\curl\B|} \sim
\frac{\sqrt{\rho\,\rho_i}}{\rho_d}\frac{\delta_i}{L}\frac{\omega_{ci}}{\omega_{cd}}\,\left(1 + \Theta\right)\,
\label{eq:comp}
\eq
where $u$ has been replaced by the \alf speed $V_A = B /\sqrt{4\,\pi\,\rho}$ and $\delta_i = V_{Ai}/\omega_{ci}$ is the ion-inertial scale. Here $L$ is the characteristic scale length of interest and $V_{Ai} = B /\sqrt{4\,\pi\,\rho_i}$ is the ion \alf\, speed. Generally $\delta_i \ll L$ and Hall term will be insignificant in comparison with the convective term. However, in the present case Hall term can become comparable to the convective term. We note that even when grains are not well coupled to the magnetic field, i.e. $\Theta \ll 1$, Hall term may still compare with the convection term. Clearly then, Hall effect is important in a weakly ionized medium. 

\section{THE PUMP WAVE}

One can investigate the dusty plasma dynamics with the help of Eqs. (\ref{eq:cont1}), (\ref{eq:meq}) and (\ref{eq:ind}) along with an isothermal equation of state. We shall assume that all physical quantities depend on $z$ only in the presence of an uniform background magnetic field $\B = (0, 0, B)$. An exact solution of Eqs. (\ref{eq:cont1}), (\ref{eq:meq}) and (\ref{eq:ind}) are finite amplitude circularly polarized \alf wave
\begin{eqnarray}
B_x = A_0\,\cos\phi\,,\,\,B_y = A_0\,\sin\phi\,,\nonumber\\
U_x = U_0\,\cos\phi\,,\,\,U_y = U_0\,\sin\phi\,.
\end{eqnarray}
Here $\phi = k_0\,z - \omega_0\,t$. The wavenumber $k_0$ and frequency $\omega_0$ of the pump wave are related by the following dispersion relation 
\bq \omega_0^2 = k_0^2\,V_A^2\,\left(1 \pm
\frac{\omega_0}{\omega_{cd}}\right)\,, 
\label{eq:pdr} 
\eq 
where $\omega_{cd} = (\rho_d/\rho)(1/ 1 + \Theta)\,Z\,e\,B/m_d\, c$ is the modified dust-cyclotron frequency. Since $\rho_d/\rho = 0.01$ in molecular clouds, irrespective of the grain magnetization, the modified dust cyclotron frequency will be much smaller than the unmodified dust cyclotron frequency. The amplitudes $A_0$ and $U_0$ of the pump waves are related by 
\bq 
U_0 = - \left(\frac{A_0\,\omega_0}{k_0\,B}\right)\left(1 \pm
\frac{\omega_0}{\omega_{cd}}\right)^{-1}\,. 
\eq 
We shall assume now that the steady-state background consists of the unperturbed as well as the circularly polarized pump waves, i.e. $\B = (B_x(z), B_y(z), B)$ and $\u = (U_x(z), U_y(z), 0)$ with a constant density. The linearization of Eqs. (\ref{eq:cont1}), (\ref{eq:meq}) and (\ref{eq:ind}) yields
\begin{eqnarray}
\delt \drho + \grad\cdot\left(\rho\,\du + \drho\, \u\right) = 0\,,
\nonumber\\
\delt \du + \left(\u + \du \cdot\grad\right)\,\left(\u + \du \right) =
- \frac{\grad \dP}{\rho} + \frac{\drho}{\rho^2}\,
\grad P + \frac{1}{c\,\rho}\left(\dJ\cross\B+J\cross\dB -
\frac{\drho}{\rho}\J\cross\B\right)\,,
\nonumber\\
\delt {\dB} = \curl\left[\du\cross\B + \u\cross\dB -
\frac{1 + \Theta}{Z\,e\,n_d}\left(\dJ\cross\B + \J\cross\dB -
\frac{\drho}{\rho}\J\cross\B \right) \right]\,.
\label{rlin}
\end{eqnarray}
The above set of Eqs. (\ref{rlin}) is supplemented with an equation of state $\dP = C_s^2\,\drho$. In what follows, we shall also assume that all the perturbed quantities depend on $z$ only. Then writing above set of equations in the component form
\begin{eqnarray}
\delt\drho + \rho\,\delz\duz = 0\,,
\nonumber\\
\delt{\dux} + \duz\,\delz {U_x}
 = \frac{B}{4\,\pi\,\rho}\,\left(\delz\dBx
- \frac{\drho}{\rho}\,\delz {B_x}\right)\,,
\nonumber\\
\delt\duy + \duz\,\delz {U_y}  = \frac{B}{4\,\pi\,\rho}\,\left(\delz\dBy
- \frac{\drho}{\rho}\delz {B_y}\right) \,,
\nonumber\\
\delt\duz = - \frac{C_s^2}{\rho} \,\delz \drho
- \frac{1}{4\,\pi\,\rho} \,\left(B_x\,\delz \dBx + B_y\,\delz \dBy\right)
+ \frac{\drho}{8\,\pi\,\rho}\,\delz{\left(B_x^2 + B_y^2\right)}\,,
\nonumber\\
\delt\dBx + \duz \delz {B_x}
 = B_z\,\delz \dux
- B_x\,\delz\duz + \alpha \left(\frac{\partial^2\dBy}{\partial z^2}
- \frac{1}{\rho}\,\delz\drho\,\delz\dBy\right)\,,
\nonumber\\
\delt\dBy + \duz \delz {B_y} = B_z\,\delz\duy
- B_y\,\delz\duz - \alpha \left(\frac{\partial^2\dBx}{\partial z^2}
- \frac{1}{\rho}\,\delz\drho\,\delz\dBx\right)\,
\label{eq:lin1}
\end{eqnarray}
where $\alpha = c\,B\,\left(1 + \Theta\right) /(4\,\pi\,Z\,e\,n_d)$. While writing above equations,
$\dBz = 0$ has been assumed. Eqs. (\ref{eq:lin1}) have periodic coefficients and belongs to a general class of equations known as Floquet equation. The Floquet theorem prescribes the form of the solution for the differential equation with periodic coefficients. However, Floquet theorem also states that there exists a linear transformation that can reduce the system of ordinary differential equations (ODEs) with periodic coefficient to a system of ODEs with constant coefficient. For a non-dispersive MHD case, such a transformation has recently been proposed \citep{rs03}. In the present case, applying same transformation, it is shown that the system of Eqs.(\ref{eq:lin1})can be reduced to ODEs with constant coefficient. Defining
\begin{eqnarray}
\dBp = \dBx\,\cos\phi + \dBy\,\sin\phi\,,\,\,\dBm = \dBx\,\sin\phi
- \dBy\,\cos\phi\,,\nonumber\\
\dup = \dux\,\cos\phi + \duy\,\sin\phi\,,\,\,\dum = \dux\,\sin\phi
- \duy\,\cos\phi\,,
\end{eqnarray}
Eqs.(\ref{eq:lin1}) can be reduced to following set of equations with constant coefficient.
\begin{eqnarray}
\delt\drho + \rho\,\delz\duz = 0\,,
\nonumber\\
\delt\dup - \omega_0\,\dum = \frac{B}{4\,\pi\,\rho}\,
\left( \delz\dBp + k_0\,\dBm\right)\,,
\nonumber\\
\delt \dum + \omega_0\,\dup - k_0\,U_0\,\duz = \frac{B}{4\,\pi\,\rho}\,
\left( \delz \dBm - k_0\,\dBp\right) + k_0 \, \frac{B\,A_0}{4\,\pi\,\rho^2}\,\drho\,,
\nonumber\\
\delt \duz = - \frac{C_s^2}{\rho} \,\delz\drho
- \frac{A_0}{4\,\pi\,\rho}\,\delz\dBp\,,
\nonumber\\
\delt\dBp - \omega_0\,\dBm = B\left( \delz\dup
+ k_0\,\dum\right)
- \alpha \left( \frac{\partial^2 \dBm}{\partial z^2}
- 2\,k_0\,\delz \dBp + k_0^2\,\dBm\right)
\nonumber \\ - A_0\,\delz\duz -
\frac{\alpha\,k_0\,A_0}{\rho}\delz\drho\,,\nonumber\\
\delt\dBm + \omega_0\,\dBp =
B\left( \delz\dum - k_0\,\dup\right)
+ \alpha \left( \frac{\partial^2 \dBp}{\partial z^2}
+ 2\,k_0\,\delz \dBm - k_0^2\,\dBp\right)
\nonumber\\
+ k_0\,A_0\,\duz\,,
\label{eq:lfin}
\end{eqnarray}
One notes that the $x$ and $y$ components of the induction Eq.(Eqs. (\ref{eq:lin1})) are symmetric with respect to the Hall terms
 (with coefficient $\alpha$)  whereas in the Eq. (\ref{eq:lfin}) such symmetry does not exist. The reason for this loss of symmetry is due to the form of $\dBp$ and $\dBm$. The terms proportional to the density fluctuation in the induction Eq. (\ref{eq:lin1}), cancels out for $\dBm$. This results in the asymmetric equations for $\dBp$ and $\dBm$.

\section{DISPERSION RELATION}

The above set of equations has constant coefficient and thus, one can Fourier analyse the spatial and temporal dependence of the fluctuations as $\sim \exp{\left(i\,\omega\,t - i\,k\,z\right)}$. Denoting $\omega = \omega / \omega_0$, $k = k/k_0$, $F_{\pm} = 1/( 1 \pm \omega/\omega_{cd})$, $\beta = C_s^2/V_A^2$ and $C_s^2 = k^2\,\beta F$ following $8^{th}$ order dispersion relation is derived from Eq. (\ref{eq:lfin}). 
\bq 
\omega^8 + a_7\, \omega^7 + \ldots + a_1\,\omega + a_0 = 0\,. 
\label{eq:dr} 
\eq 
The coefficients $a_7,\ldots a_0$ are given in the appendix. The transition from stability to instability will proceed through $\omega = 0$. In the $\omega \rightarrow 0$ limit, 
\bq 
\omega = - \frac{a_0}{a_1}\,. 
\label{eq:ldr} 
\eq 
In the long wavelength limit retaining only $\sim O(k),\, O(k^2)$ terms in the coefficients $a_1$ and $a_0$ one may write 
\bq 
\frac{a_1}{k\,F_{\pm}} = - 2 \frac{A_0^2}{B^2} \left[ 1 -
F_{\pm} \left( 1 + \frac{\omega_0}{\omega_{cd}} \right) \right] +
k\,\frac{A_0^2}{B^2}\left( 1 + F_{\pm} \right), 
\eq
\bq 
\frac{a_0}{k\,F_{\pm}} = \left[\beta\,\left( 1 + F_{\pm}
\frac{\omega_0}{\omega_{cd}} \right) -  F_{\pm} \frac{A_0^2}{B^2}
\right] \left[ 1 - F_{\pm} \left( 1 +
\frac{\omega_0}{\omega_{cd}}\right) \right] 
\label{eq:caz} 
\eq
Recall that for the left-circularly polarized pump waves, $F_{+} = 1/(1 + \omega_0/\omega_{cd})$ and hence, from Eq. (\ref{eq:ldr}), $a_0 = 0$. Thus in view of Eq. (\ref{eq:caz}), one should anticipate that the instability will disappear in the vicinity of $k = 0$. As will be shown later, the numerical solution of full dispersion relation, Eq. (\ref{eq:dr}) indeed supports this conclusion.

For the right-circularly polarized pump waves, when $F_{-} = 1/(1 -\omega_0/\omega_{cd})$, the growth rate can be written as 
\bq
Im[\omega] = \frac{F_{-}} {2}\, \frac{A_0^2}{B^2} \left( \beta -
\frac{A_0^2}{B^2}\right)\,. 
\label{eq:lres} 
\eq 
As is clear from above Eq. (\ref{eq:lres}), the growth rate is inversely proportional to the factor $(1- \omega_0/\omega_{cd})$. This implies that near
$\omega_0 \simeq \omega_{cd}$, when pump is operating near the modified dust-cyclotron frequency, the instability can grow resonantly. Therefore, the growth of the parametric instability is quite different for the left and right-circularly polarized pump. Whereas for the left-circularly polarized waves, the instability does not exist in the neighborhood of $k = 0$, for the right-circularly polarized pump, the instability may become large near $k = 0$. Clearly, for the right circularly polarized pump, unbounded growth of the instability is not possible. The linear approximation in which the dispersion relation (\ref{eq:dr}) has been derived breaks down for any dependent physical variable $f$ once $\delta f \leq f$. Therefore, in the vicinity of resonance nonlinear terms needs to be considered. This task is beyond the scope of the present work.

Now the dispersion relation (\ref{eq:dr}) is solved numerically. In
Fig. 1(a) the normalized growth rate is plotted against the
normalized wave number for different values of
$\omega_0/\omega_{cd}$. The left circularly polarized pump is
assumed. The instability does not exist in the vicinity of $k
\rightarrow 0$ and $k = 0.2$ is the threshold of the instability. It
is in conformity with the above discussion where in the vicinity of
$k = 0$, wave does not grow. This is caused by the dissipation of
the long wavelength fluctuations by the plasma-neutral collisions.
With the increase in $\omega_0/\omega_{cd}$, the growth rate
decreases. As is clear from the pump wave dispersion relation
(\ref{eq:pdr}), the increase in $\omega_0/\omega_{cd}$ implies the
increasing importance of the Hall term ($\J\times\B$), which appears
due to the relative drift between the plasma and the neutral. This
drift is caused entirely by the collisonal momentum exchange.
Therefore, the increase in $\omega_0/\omega_{cd}$ implies the
increased dissipation of the free energy. Hence with the increasing
$\omega_0/\omega_{cd}$, one would expect a decrease in the growth
rate. With decreasing $\omega_0/\omega_{cd}$, the growth rate is due
to nondispersive \alf pump and will correspond to a non-dissipative,
ideal regime. Hence one sees the saturation of the growth rate with
decreasing with decreasing $\omega_0/\omega_{cd}$.

In Fig. 1(b) the growth rate of the parametric instability is shown
against the variation of the pump amplitude $A_0/B$. Recall that $B$
is the amplitude of the constant magnetic field in the $z$ direction
and, $A_0$ is the amplitude of the transverse $x - y$ fields. The
growth rate is not very sensitive towards smaller wavelength, i.e.
after $k/k_0 \geq 0.5$. However, the onset of instability is
slightly delayed when $A_0/B = 1$ than when $A_0/B = 0.1$ implying
that the effect of collision is less pronounced for a stronger $B_z$
field.

In Fig. 2 both real and imaginary frequencies are shown for $A_0/B = 1$ and $\omega_0/\omega_{cd} = 1$. We see that from the onset until the demise of the instability, the real part of the frequency is constant. The part of free energy that is responsible for setting the normal oscillation in the system has been diverted towards the wave growth. If one superposes the imaginary part on the real frequency, then one gets an almost constant $\omega_r$. This should be expected since the total energy available to the system is due to the \alf pump, and thus, the growth of the fluctuation occurs at the expense of the oscillatory real part.

The growth rate is not sensitive to the plasma $\beta$. Unlike in
the non-dispersive, non-dissipative case where the increase of
$\beta$ leads to the increase in the growth rate \citep{g78},
in the present dissipative, dispersive case, plasma $\beta$ appears
to have no role. In the non-dissipative case the increase in $\beta$
implies the unlikelihood of compressional wave excitation, the role
of plasma $\beta$ in the dissipative case is unclear. The dispersive
nature of the pump wave is due to the collisonal effects and thus,
$\beta$ is not directly related to the growth of the waves.

In Fig.3(a) the growth rate and in Fig. 3(b) corresponding real part
of the frequency is shown for the right-circularly polarized pump.
When $\omega_0/\omega_{cd} = 0.9$, the growth rate becomes very
large. The free energy is resonantly pumped into the fluctuations
with increasing $\omega_0/\omega_{cd}$. The physical system behaves
like a driven oscillator. The resonant driving is indirectly related
to the neutral-plasma collisions. The relative drift between the
plasma and the dust causes a Hall field over the dust-cyclotron
time. If the \alf wave propagation time $\omega_0^{-1}$ becomes
comparable to the dust-cyclotron time, the energy is freely fed to
the fluctuation by the  pump to the plasma particles. Resulting free
energy to the fluctuation causes the large growth rate. This
behavior has been analytically predicted in the limiting case by Eq.
(\ref{eq:lres}). It should be noted from the corresponding curve in
Fig. 3(b) that the real frequency shows a sharp decline for
$\omega_0/\omega_{cd} = 0.9$. This suggest that almost all the free
pump energy is resonantly used in the fluctuation growth. Similar
behavior is also noted for $\omega_0/\omega_{cd}= 0.8$ and
$\omega_0/\omega_{cd}= 0.01$. Since the instability growth rate in
this case is smaller than when $\omega_0/\omega_{cd} = 0.9$, the
part of the available free energy remains in the real part of the
frequency. This is in agreement with the well known behavior of the
oscillators near resonance although present system is more complex.

\section{APPLICATION AND SUMMARY}

The grains are important charge carriers in the molecular clouds and their presence can significantly alter the dynamics of the low frequency \alf wave in the neutral medium. Observations of molecular clouds suggest a magnetically threaded supersonically turbulent environment, and the generation of magnetohydrodynamic waves in such an environment is inevitable. Thus taking the upper value of the pump wave frequency \citep{fhw04}, $\omega_0 \simeq10^{-4}\,\mbox{yr}^{-1}$, the growth rate of the left-circularly polarized wave $0.3\, \omega_0$ suggests that the parametric instability of the \alf wave could be relevant to the onset of turbulence. The \alf wavelength lies in the range of $0.07- 0.35 \mbox{pc}$(Folini et al. 2004) and since the maximum growth rate occurs at $k/k_0 = 0.5$ (Fig. 1(a)), the parametric instability of the left-circularly polarized mode will operate between $0.1 - 0.7\,\mbox{pc}$.

To check whether the resonance condition can prevail in the molecular clouds, we shall assume that the grains are magnetized. Since smaller grains are more likely coupled to the magnetic field than the large grains, we take $m_g = 10^{-18} g$. For a typical magnetic field $B \sim 10 \mu\mbox{G}$, and taking $\rho_d/\rho = 0.01$ and $\Theta = 10$, one gets $\omega_{cd} \sim 10^{-4}\,\mbox{yr}^{-1}$ which is comparable to the pump frequency. Therefore, the right circularly polarized wave can resonantly excite this instability. Furthermore, it can operate on very long wavelengths.

To summarize, parametric instability of a weakly ionized medium may
play an important role in exciting the turbulence in the
interstellar medium. Following is the itemized summary of the
present work.

1. The parametric instability of the weakly ionized collisonal
medium is studied in the present work. It is shown that the
collision causes the dispersive nature of the pump waves.

2. Using a linear transformation, the linearized equations with
periodic coefficients are reduced to a form with constant
coefficient allowing the validity of the normal mode analysis at all
wavelengths.

3. The instability is sensitive to the change in the ratio of the
pump and the cyclotron frequencies and weakly sensitive to the ratio
$A_0/B$. The growth rate is not sensitive to the plasma $\beta$.

4. The instability can become an order of magnitude larger for the
right hand circularly polarized pump particularly near the
resonance, i.e. when $\omega_0 \simeq \omega_{cd}$.
\\

\acknowledgments
The encouragement of an unknown referee in improving the manuscript is gratefully acknowledged. Support of the Australian Research Council is gratefully acknowledged.

\appendix

\section{Appendix material}

{\bf{Coefficient of the dispersion relation (\ref{eq:dr})}}

\begin{eqnarray}
a_7 =  4\,k\,F_{\pm}\,,\quad
\frac{a_6}{F_{\pm}} = - \left[ 1 + 2 \frac{\omega_0}{\omega_{cd}}
+ k^2\,\frac{\omega_0}{\omega_{cd}} \right]
\left[ 1 +  k^2\,\frac{\omega_0}{\omega_{cd}}\right]\nonumber\\
-k^2 - 2\,F_{\pm}^{-1}
- \beta\,k^2 - k^2\,\left(\frac{A_0}{B}\right)^2 + 4\,k^2\,F_{\pm}\,
\left(\frac{\omega_0}{\omega_{cd}}\right)^2
- \left(1+k^2\right)\,.
\end{eqnarray}
\begin{eqnarray}
\frac{a_5}{k\,F_{\pm}} =
\left(1 + \left(\frac{A_0}{B}\right)^2\right)\,\left( 1 -
\frac{\omega_0}{\omega_{cd}}\,\left(1 - k^2\right)\,F_{\pm}\right)
\nonumber\\
+ 2\,\left( 1 + \frac{\omega_0}{\omega_{cd}}\,\left(1 +
k^2\right)\,F_{\pm}\right)
- \left(\frac{A_0}{B}\right)^2 + 1
- 4\,\frac{\omega_0}{\omega_{cd}} - 2\,\frac{\omega_0}
{\omega_{cd}}\,k^2\,\beta\,F_{\pm}
\nonumber\\
- \left(\frac{A_0}{B}\right)^2\,F_{\pm}\,k^2\,
\frac{\omega_0}{\omega_{cd}}
-2\,\frac{\omega_0}{\omega_{cd}}\,\left[k^2\,
F_{\pm}+1+k^2\,\beta\,F_{\pm}+\left(\frac{A_0}{B}\right)^2\,
F_{\pm}\,k^2\right]
\nonumber\\
- 2\,\frac{\omega_0}{\omega_{cd}}\,\left[1 + F_{\pm}\,
\left(1 + k^2\right)\right]
+ k^{-1}\,\left(1 + k\right)
\end{eqnarray}
\begin{eqnarray}
\frac{a_4}{k\,F_{\pm}} = k\,\beta + k\left(\frac{A_0}{B}\right)^2
+ k\,\left[F_{\pm}\,\left(\frac{A_0}{B}\right)^2 + \left(k^2\,
\beta\,F_{\pm} + \left(\frac{A_0}{B}\right)^2\right)\right]
\nonumber\\
+ 2\,k\,F_{\pm}\,\frac{\omega_0}{\omega_{cd}}
\left[-\left(\frac{A_0}{B}\right)^2
+ 1 - 2\,\frac{\omega_0}{\omega_{cd}} - 2\,k^2\,\beta\,
F_{\pm}\frac{\omega_0}{\omega_{cd}}
- k^2\,F_{\pm}\frac{\omega_0}{\omega_{cd}}
\left(\frac{A_0}{B}\right)^2\right]
\nonumber\\
+ k\,\left[1 + k^{-2}\,F_{\pm}^{-1} + \beta +
\left(\frac{A_0}{B}\right)^2\right]\,
\left[1 + F_{\pm}\,\left(1 + k^2\right)\right]
- 2\,\frac{\omega_0}{\omega_{cd}}\,F_{\pm}\left[2\,k\,
\frac{\omega_0}{\omega_{cd}}- 1 - k\right]
\nonumber\\
+ k\,\left[\left(\frac{A_0}{B}\right)^2  +
 k^{-2}\,\frac{\omega_0}{\omega_{cd}}\,
\left(1 + k^2\right) + \beta\,\left( 1 + F_{\pm}\,
\frac{\omega_0}{\omega_{cd}}\,\left(1 + k^2\right)\right)
\right. \nonumber\\ \left.
- 1 + k^{-2}\,F_{\pm}^{-1} \right]
\,\left( 1 - \frac{\omega_0}{\omega_{cd}}\,
\left(1 - k^2\right)\,F_{\pm}\right)
- 2\,k\,F_{\pm} \left[1 + \left(\frac{A_0}{B}\right)^2\right]
\nonumber\\
+ \left( 1 + \frac{\omega_0}{\omega_{cd}}\,
\left(1 + k^2\right)\,F_{\pm}\right)\,
\left[k^{-1}\,F_{\pm}^{-1} - \frac{\omega_0}{\omega_{cd}}\,
k^{-1}\left(1 - k^2\right) - k^{-1}\,\left(1 + k^2\right)\right]
\end{eqnarray}
\newpage
\begin{eqnarray}
\frac{a_3}{k\,F_{\pm}} = - \left[\left(\frac{A_0}{B}\right)^2
+ k^2\,F_{\pm}\left(\beta + \left(\frac{A_0}{B}\right)^2\right)
+ \left(\frac{A_0}{B}\right)^2\right]\,\left( 1 - \frac{\omega_0}
{\omega_{cd}}\,\left(1 - k^2\right)\,F_{\pm}\right)
\nonumber\\
- 2\,k\,F_{\pm}\left[k\left(\left(\frac{A_0}{B}\right)^2 - 1\right)
+ \frac{1}{k\,F_{\pm}}
+ k\,\beta \left( 1 + \frac{\omega_0}{\omega_{cd}}\,
\left(1 + k^2\right)\,F_{\pm}\right)
\right.
\nonumber\\
\left.
+ \frac{\omega_0}{\omega_{cd}}\frac{\left(1 + k^2\right)}{k}\right]
- \left[1 + \left(\frac{A_0}{B}\right)^2\right]
\left[1 - \frac{\omega_0}{\omega_{cd}}\,F_{\pm}\,\left(1 - k^2\right)
- F_{\pm}\,\left(1 + k^2\right)\right]
\nonumber\\
+ 2\,k^2\,F_{\pm}\,\frac{\omega_0}{\omega_{cd}} \left(\beta
+ \left(\frac{A_0}{B}\right)^2\right)
- \left[k^2\,\beta\,F_{\pm}\, + \left(\frac{A_0}{B}\right)^2\,F_{\pm}\,k^2\right]
\nonumber\\
+ 2\,k\,F_{\pm}\,\frac{\omega_0}{\omega_{cd}}\,\left[k
\left( \beta + \left(\frac{A_0}{B}\right)^2\right) +
k\,\left(F_{\pm}\,\left(\frac{A_0}{B}\right)^2
+ k\,\left(k^2\,\beta\,F_{\pm}\, +
\left(\frac{A_0}{B}\right)^2\right)\right)\right]
\nonumber\\
+ \left[\left(\frac{A_0}{B}\right)^2 - 1 + 2\,
\frac{\omega_0}{\omega_{cd}}
+ 2\,\frac{\omega_0}{\omega_{cd}}\,k^2\,\beta\,F_{\pm}\,
+ \left(\frac{A_0}{B}\right)^2\,F_{\pm}\,k\,\frac{\omega_0}{\omega_{cd}}\right]\,
\left[1 + F_{\pm}\,\left(1 + k^2\right)\right]
\nonumber\\
+ \left[k^2\,F_{\pm} + 1 + k^2\,\beta\,F_{\pm}\, +
\left(\frac{A_0}{B}\right)^2\,F_{\pm}\,k^2\right]\,\left[2\,k\,F_{\pm}\,
\frac{\omega_0}{\omega_{cd}} - F_{\pm}\,\left(1 + k\right)\right]
\end{eqnarray}
\begin{eqnarray}
\frac{a_2}{k\,F_{\pm}} = -\left[k\,\beta\left( 1 + \frac{\omega_0}{\omega_{cd}}\,
\left(1 + k^2\right)\,F_{\pm}\right)
- k\,F_{\pm}\left(k^2\,\beta + \left(\frac{A_0}{B}\right)^2\right)\right]
\nonumber\\
\left( 1 - \frac{\omega_0}{\omega_{cd}}\,\left(1 - k^2\right)\,F_{\pm}\right)
+ 2\,k\,F_{\pm}\left[2\left(\frac{A_0}{B}\right)^2 + k^2\,F_{\pm}\left(\beta
+ \left(\frac{A_0}{B}\right)^2\right)\right]
\nonumber\\
+ \left[k\,(\left(\frac{A_0}{B}\right)^2 - 1) + \frac{1}{k\,F_{\pm}}
+ \frac{\omega_0}{\omega_{cd}}\,F_{\pm}\,
\frac{1 + k^2}{k}
+ k\,\beta\left( 1 + \frac{\omega_0}{\omega_{cd}}\,\left(1 + k^2\right)
\,F_{\pm}\right)\right]
\nonumber\\
\,\left[-1 + \frac{\omega_0}{\omega_{cd}}\,F_{\pm}\,\left(1 - k^2\right)
+ F_{\pm}\,\left(1 + k^2\right)\right]
\nonumber\\
+ 2\,k\,F_{\pm}\,\left[2\,k\,\beta\frac{\omega_0}{\omega_{cd}}
+ k\frac{\omega_0}{\omega_{cd}} \left(\frac{A_0}{B}\right)^2 - k^2\,
F_{\pm}\,\frac{\omega_0}{\omega_{cd}}\,
\left(\beta + \left(\frac{A_0}{B}\right)^2\right)\right]
\nonumber\\
- \left[k\,\beta + k\,\left(\frac{A_0}{B}\right)^2 + k\,F_{\pm}\,
\left(\frac{A_0}{B}\right)^2
+ k\,\left(k^2\,\beta\,F_{\pm}\, + \left(\frac{A_0}{B}\right)^2\right)\right]
\nonumber\\
\,\left[1 + F_{\pm}\,\left(1 + k^2\right)\right]
+ F_{\pm}\,\left[\left(\frac{A_0}{B}\right)^2  - 1 +
\frac{\omega_0}{\omega_{cd}}
+ 2\,k^2\,F_{\pm}\frac{\omega_0}{\omega_{cd}}\left(2\,\beta
+ \left(\frac{A_0}{B}\right)^2\right)\right]\,
\nonumber\\
\left[2\,k\frac{\omega_0}{\omega_{cd}} - 1 - k\right]
\end{eqnarray}
\newpage
\begin{eqnarray}
\frac{a_1}{k\,F_{\pm}} = 2\,k^2\,\beta\,F_{\pm}\,
\left[1 + F_{\pm}\,\left(\frac{\omega_0}{\omega_{cd}} +
k^2\,\left(\frac{\omega_0}{\omega_{cd}} - 1 \right)\right)
- \frac{F_{\pm}\,A_0^2}{\beta\,B^2}
\right]
\nonumber\\
- \left(\frac{A_0}{B}\right)^2\,
\left[2 + F_{\pm}\,k^2
\left( 1 + \beta\left(\frac{B}{A_0}\right)^2 \right)\right]
\,\left[1 - F_{\pm}\,\left(\frac{\omega_0}{\omega_{cd}}
\left(1 - k^2\right) + 1 + k^2\right)\right]
\nonumber\\
- k^2\,F_{\pm}\,\left[\beta\,\left(2\,
\frac{\omega_0}{\omega_{cd}} - 1\right)
+ \left(\frac{A_0}{B}\right)^2\,\left(
\frac{\omega_0}{\omega_{cd}} - 1\right)\right]
\,\left[1 + F_{\pm}\,\left(1 + k^2\right)\right]
\nonumber\\
- k^2\,F_{\pm}\,\left[\beta\,\left(1 + k^2\,F_{\pm}\right) +
 \left(\frac{A_0}{B}\right)^2\,
\left(2 + F_{\pm}\right)\right]
\left[2\,\frac{\omega_0}{\omega_{cd}} - \frac{1+k^2}{k}\right]\,.
\end{eqnarray}
\begin{eqnarray}
\frac{a_0}{k^2\,F_{\pm}} = k\,F_{\pm}^2\left[\left( 2\beta
+ \frac{A_0^2}{B^2}\right)\frac{\omega_0}{\omega_{cd}} -
\left(\beta + \frac{A_0^2}{B^2}\right)\right]\,
\left[-2\,F_{\pm}\frac{\omega_0}{\omega_{cd}} + 1 + k\right]+
\nonumber\\ \left[\beta
\,F_{\pm}\frac{\omega_0}{\omega_{cd}}\left(1 + k^2\right)
+ \beta - F_{\pm}\,\left(\beta\,k^2
+ \frac{A_0^2}{B^2}\right)\right]\,
\left[\left( 1 - \frac{\omega_0}{\omega_{cd}}\,\left(1 -
k^2\right)\,F_{\pm}\right) - F_{\pm}\,\left(1 + k^2\right)\right].
\label{coef}
\end{eqnarray}

\clearpage

\begin{figure}
\epsscale{.80}
\plotone{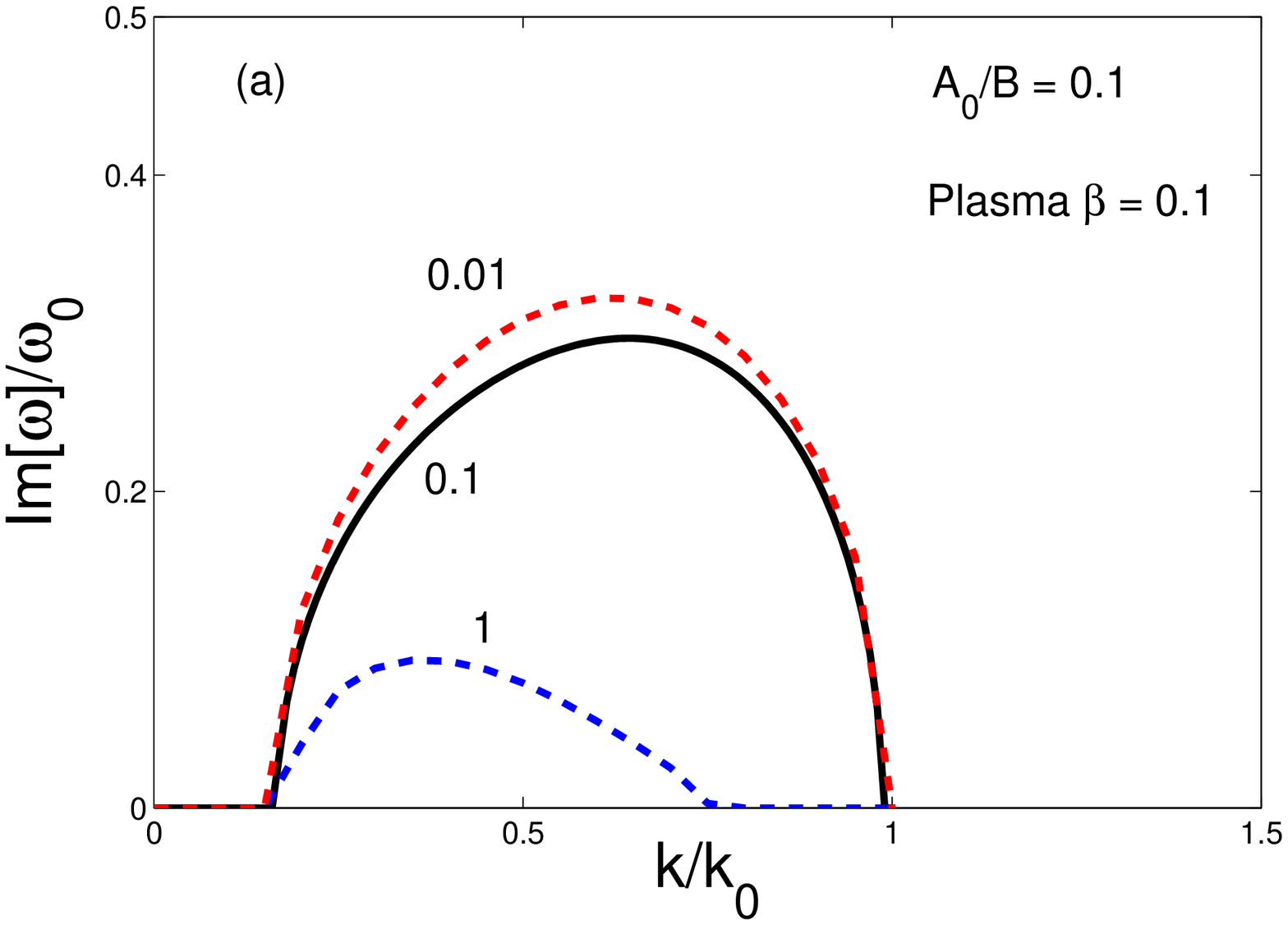}
\plotone{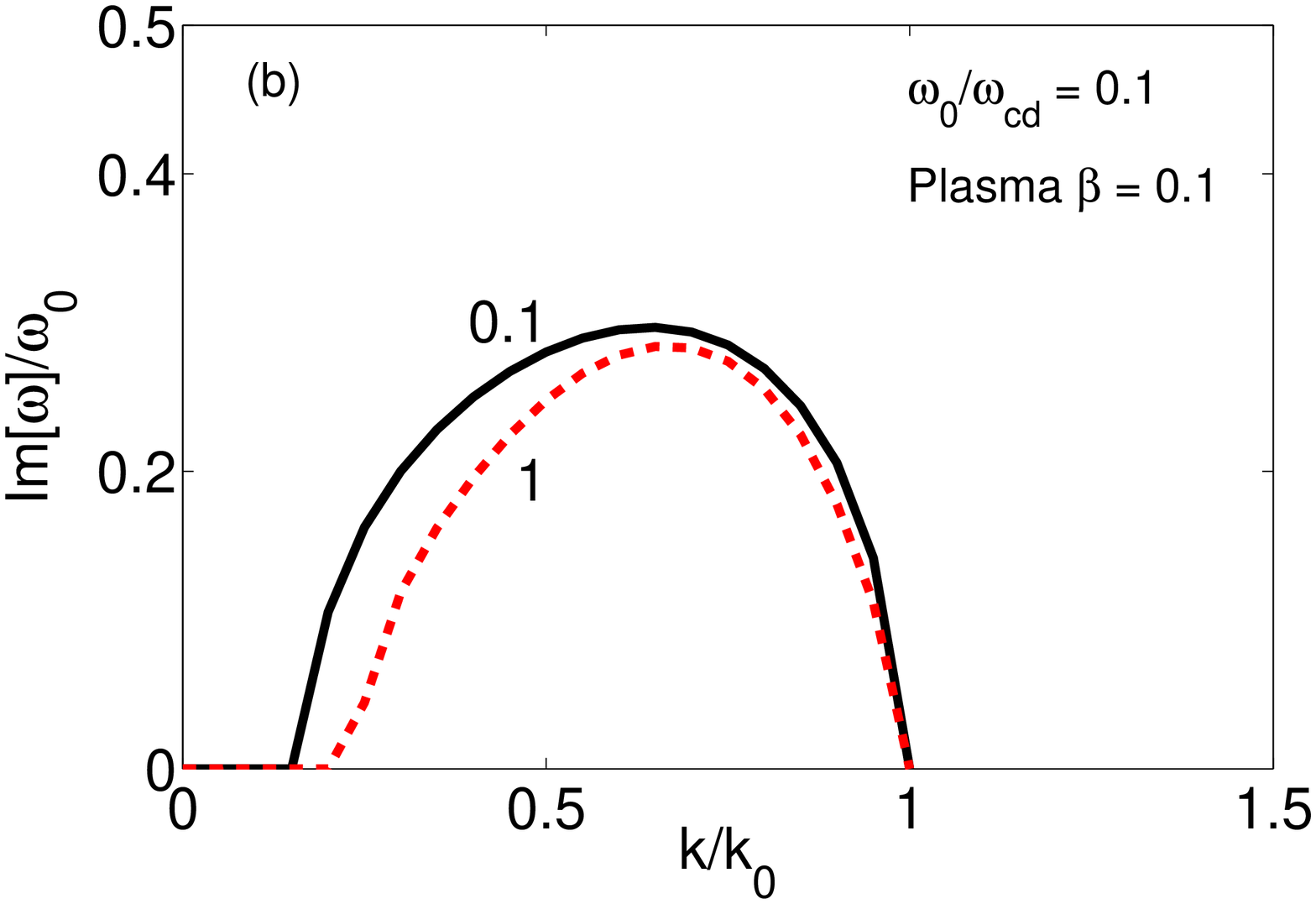}
\caption{The dependence of the growth rate on the ratio of the left
circularly polarized pump wave to the dust-cyclotron frequency is
shown in Fig. 1(a) for the corresponding physical parameters in the
box. In Fig. 1(b) the dependence of the growth rate on the amplitude
of the pump wave $A_0/B$ is shown.}
\end{figure}

\clearpage

\begin{figure}
\epsscale{.80}
\plotone{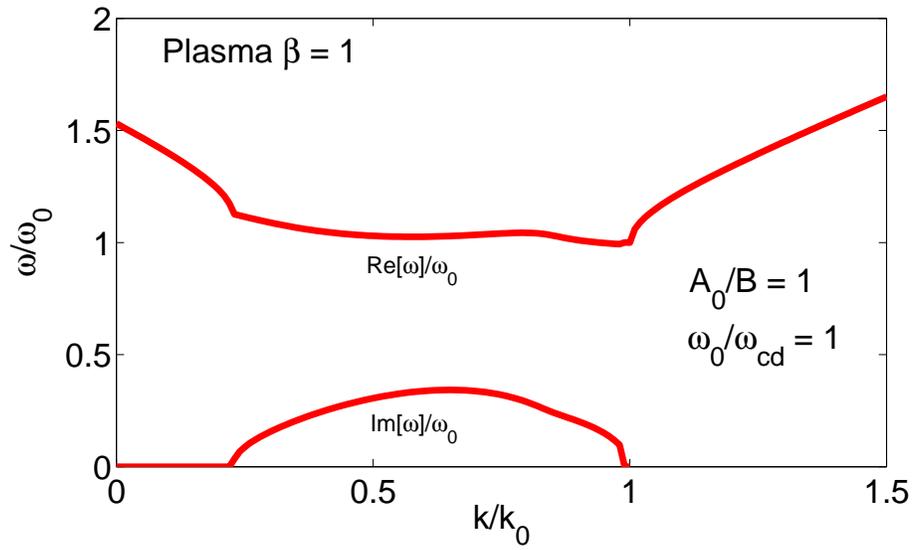}
\caption{ Both  real  and  imaginary  part of $\omega/\omega_{0}$ 
is plotted against $k/k_0$ for the plasma $\beta$, $\omega_0/\omega_{cd}$ and  
$A_0/B$ shown in the box.}
\end{figure}

\clearpage

\begin{figure}
\epsscale{.80}
\plotone{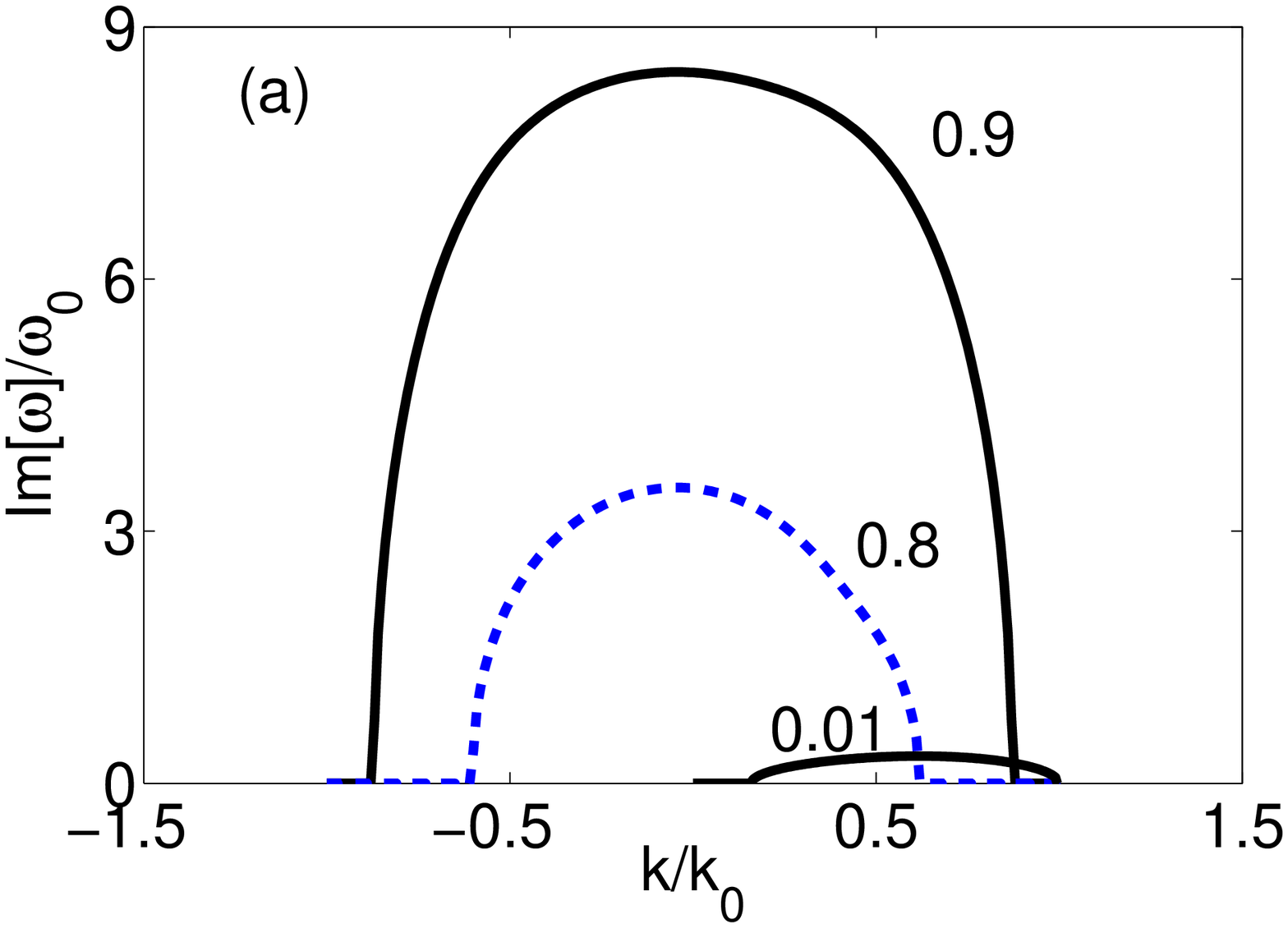} 
 \plotone{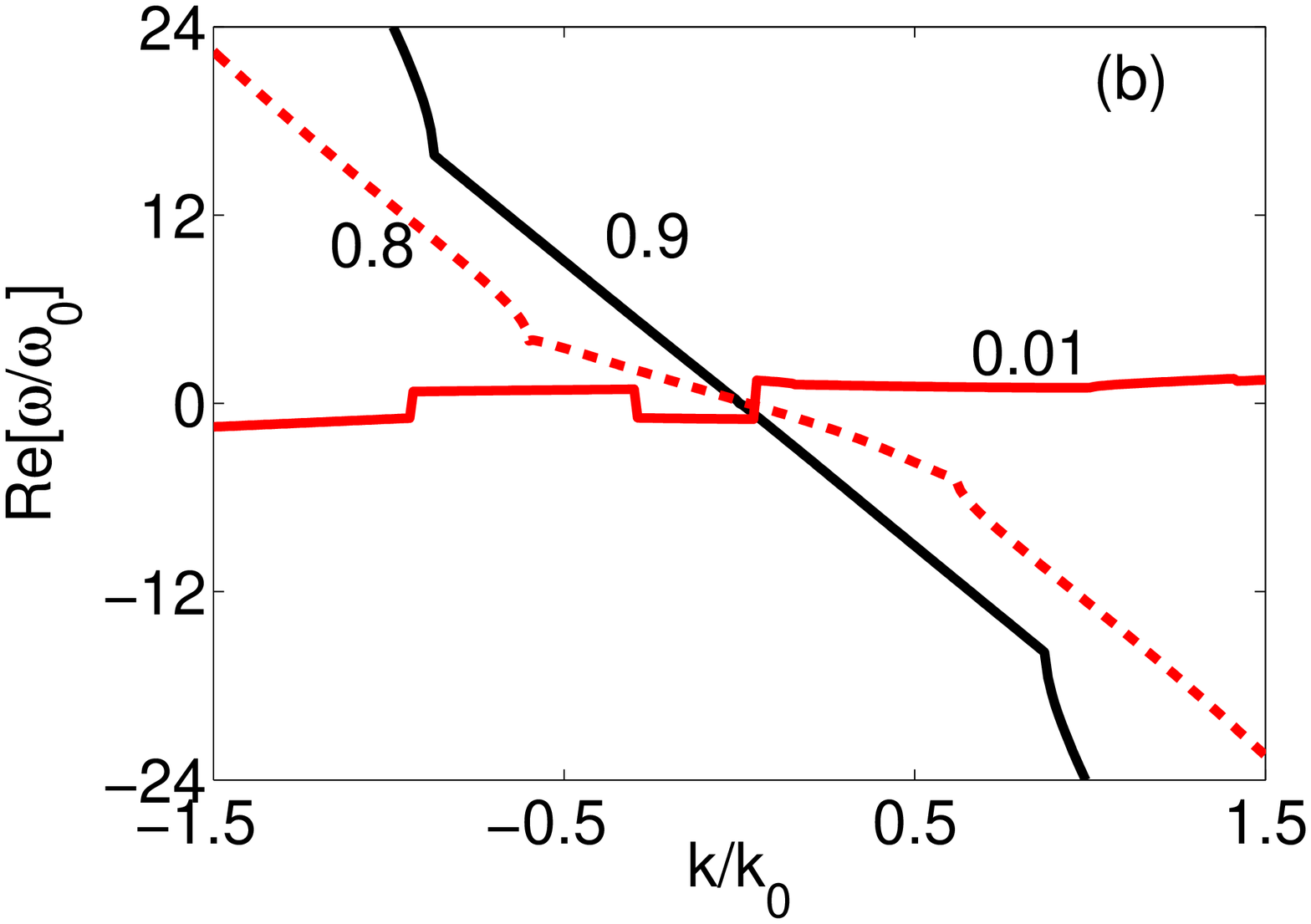}
\caption{The dependence of the growth rate on the amplitude of the
right handed circularly polarized pump wave is shown in Fig. 3(a)
and corresponding real part of the frequency is shown in Fig. 3(b)
for different values of $\omega/\omega_{cd}$. The value of plasma
$\beta$ and $A_0/B$ is $0.1$.}
\end{figure}

\end{document}